\documentclass[pra,showpacs,twocolumn,eqsecnum,superscriptaddress]{revtex4}

\usepackage{makeidx}
\usepackage{eurosym} 
\usepackage{amssymb}
\usepackage{amsfonts}
\usepackage{amsmath}
\usepackage{setspace}
\usepackage[english]{babel}
\usepackage{graphicx}
\usepackage[applemac]{inputenc}
\usepackage{color}
\usepackage{ifthen}
\usepackage{float}
\usepackage{verbatim}

\begin{document}

\newcommand{\leftexp}[2]{{\vphantom{#2}}^{#1}{#2}}
\newcommand{\leftind}[2]{{\vphantom{#2}}_{#1}{#2}}
\newcommand{\bra}[1]{\langle #1|}
\newcommand{\ket}[1]{|#1\rangle}
\newcommand{\braket}[2]{\langle #1|#2\rangle}
\newcommand{\wt}[1]{\widetilde{#1}}
\newcommand{\dblebar}[1]{\overline{\overline{#1}}}
\newcommand{\axe}[1]{\textbf{#1}}
\newcommand{\mat}[1]{\textbf{#1}}
\newcommand{\dvpart}[2]{\frac{\partial #1}{\partial #2}}
\newcommand{\dvtot}[2]{\frac{d #1}{d #2}}
\newcommand{\ddvpart}[2]{\frac{\partial^2 #1}{\partial #2 ^2}}
\newcommand{\ddvtot}[2]{\frac{d^2 #1}{d #2 ^2}}
\newcommand{\dg}{^\dagger}

\def\be{\begin{equation}}
\def\ee{\end{equation}}

\def\bs{\boldsymbol}
\def\bsplit{\begin{split}}
\def\nsplit{\end{split}}
\def\etal{{\it et al. }}
\def\fdsl{\frac{d}{\ell}}
\def\dsl{d/\ell}
\def\ecsl{\left(e^2/\epsilon \ell\right)}
\def\fecsl{\left(\frac{e^2}{\epsilon \ell}\right)}
\def\ua{\uparrow}
\def\da{\downarrow}
\def\cpt{\text{CP}^{3}}
\def\exp{\text{exp}}
\def\dpspn{\frac{2\pi}{\Phi_0}}

\newcommand{\eq}[1]{(\ref{#1})}

\newcommand{\red}{\color[rgb]{0.8,0,0}}
\newcommand{\green}{\color[rgb]{0.0,0.6,0.0}}
\newcommand{\blue}{\color[rgb]{0.0,0.0,0.6}}

\title{Ultrastrong coupling regime of cavity QED with phase biased flux qubits} 

\author{J. Bourassa}
\affiliation{D\'epartement de Physique et Regroupement Qu\'eb\'ecois sur les Mat\'eriaux de Pointe (RQMP), Universit\'e de Sherbrooke, Sherbrooke, Qu\'ebec, Canada, J1K 2R1}
\author{J. M. Gambetta}
\affiliation{Department of Physics and Astronomy and Institute for Quantum Computing, University of Waterloo, Waterloo, Ontario, Canada, N2L 3G1}
\author{A. A. Abdumalikov~Jr.}
\affiliation{The Institute of Physical and Chemical Research (RIKEN), Wako, Saitama 351-0198, Japan}
\author{O. Astafiev}
\affiliation{The Institute of Physical and Chemical Research (RIKEN), Wako, Saitama 351-0198, Japan}
\author{Y. Nakamura}
\affiliation{The Institute of Physical and Chemical Research (RIKEN), Wako, Saitama 351-0198, Japan}
\affiliation{NEC Nano Electronics Research Laboratories, Tsukuba, Ibaraki 305-8501, Japan}
\author{A. Blais}
\affiliation{D\'epartement de Physique et Regroupement Qu\'eb\'ecois sur les Mat\'eriaux de Pointe (RQMP), Universit\'e de Sherbrooke, Sherbrooke, Qu\'ebec, Canada, J1K 2R1}

\begin{abstract}
We theoretically study a circuit QED architecture based on a superconducting flux qubit directly coupled to the center conductor of a coplanar waveguide transmission-line resonator.  As already shown experimentally [Abdumalikov \etal Phys. Rev. B {\bf 78}, 180502 (2008)], the strong coupling regime of cavity QED can readily be achieved by optimizing the local inductance of the resonator in the vicinity of the qubit.  In addition to yielding stronger coupling with respect to other proposals for flux qubit based circuit QED, this approach leads to a qubit-resonator coupling strength $g$ which does not scale as the area of the qubit but is proportional to the total inductance shared between the resonator and the qubit. Strong coupling can thus be attained while still minimizing sensitivity to flux noise. Finally, we show that by taking advantage of the the large kinetic inductance of a Josephson junction in the center conductor of the resonator can lead to coupling energies of several tens of percent of the resonator frequency, reaching the ultrastrong coupling regime of cavity QED where the rotating-wave approximation breaks down. This should allow an on-chip implementation of the $E \otimes \beta$ Jahn-Teller model.
\end{abstract}


\maketitle

\section{Introduction}

Combined with the large electric dipole moment of superconducting charge qubit,  the large vacuum electric field of microwave transmision-line resonators can be used to reach the strong coupling regime of cavity QED~\cite{schoelkopf:2008a}. However, charge qubits suffer from charge fluctuations which leads to low coherence times. By working with a Cooper Pair Box qubit in a parameter regime where charge dispersion is small, the transmon qubit~\cite{koch:2007a} has led to significant improvement in coherence times~\cite{schreier:2008a}, in addition to larger qubit-field coupling strengths $g$.  This is, however, done at the cost of lower anharmonicity, limiting gate speed. In Refs.~\cite{yang:2005a,lindstrom:2007a}, it was suggested that an alternative approach to reaching the strong coupling regime with superconducting qubits is to inductively couple flux qubits to the zero-point motion magnetic field of a transmission-line resonator.  In this case, coupling increases with qubit loop area $A$, with $A\sim8\;\mu$m$^2$  expected to be sufficient to reach coupling strengths of a few tens of MHz~\cite{lindstrom:2007a}.  While comfortably in the strong coupling limit, the predicted values are almost an order of magnitude lower than what can be obtained with transmon qubits~\cite{schoelkopf:2008a}. Larger couplings can be obtained by increasing the qubit area, but only at the expense of increased sensitivity to flux noise.

In this paper, we theoretically investigate an approach experimentally realized by Abdumalikov \etal ~\cite{abdufarrukh-a.-abdumalikov:2008a} where flux qubits are directly connected to the center conductor of a coplanar waveguide transmission-line resonator.  By changing the width of the center conductor to take advantage of the kinetic inductance, the phase bias of the qubit by the resonator is enhanced.  We show how this approach leads to significant qubit-resonator coupling, easily reaching the strong-coupling regime.  Inserting a Josephson junction in the center conductor of the resonator, much stronger couplings can be obtained, with $g$ reaching several tens of percent of the resonator frequency.  In this ultra-strong coupling regime, the ubiquitous rotating-wave approximation is expected to break down, leading to as of yet unexplored physics in cavity QED.  In addition to the larger coupling, an advantage of this approach over that presented in Ref.~\cite{yang:2005a,lindstrom:2007a} is that $g$ does not scale with the qubit area. Moreover, with its multi-level structure, the flux qubit can be used in the $\Delta$ configuration~\cite{liu:2005a} opening the possibility to realize EIT in a cavity \cite{werner:1999a} and a wealth of other quantum optics phenomena circuit QED.

This paper is organized as follows. We start by finding the normal modes of an inhomogeneous transmission-line resonator. The case of a Josephson junction playing the role of the inhomogeneity is then discussed.  Building on these results, we obtain the Hamiltonian for a flux qubit directly connected to the center conductor of the inhomogeneous transmission-line and obtain expressions for the qubit-resonator coupling strength.  Finally, numerical results for the coupling strength are presented.

\section{Enhanced phase biasing}

\begin{figure}[t]
\centering
\includegraphics[width=0.45\textwidth]{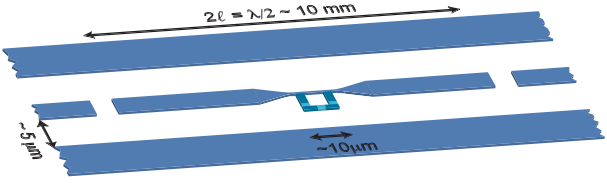}
\caption[]{(Color online) 
Schematics of a 3-junction flux qubit directly connected to the center conductor of an inhomogeneous superconducting transmission-line resonator.}
\label{fig:cavite-qubit}
\end{figure}

A schematic of the circuit we consider is shown in Fig.~\ref{fig:cavite-qubit}.  A superconducting flux qubit is fabricated such that its loop is closed by the center conductor of a transmission-line resonator of length $2\ell$ (ranging from $x=-\ell$ to $+\ell$).  Assuming a non-uniform resonator, the Lagrangian density reads
\be
\label{eqn:lag-density}
\mathcal L_\mathrm{tl}= \frac{C^0(x) \dot \psi^2(x,t)}{2} - \frac{1}{2 L^0(x)} \left(\dvpart{\psi(x,t)}{x}\right)^2, 
\ee
with $\psi(x,t) = \int_{-\infty}^t dt' V(x,t')$, $C^0(x)$ the position-dependent capacitance per unit length and $L^0(x)= L^0_{\text{geo}}(x) + L^0_{\text{kinetic}}(x)$ the position-dependent inductance per unit length including both geometrical and kinetic contributions.  

The corresponding Euler-Lagrange equation of motion
\be
\label{eqn:euler-lagrange}
\dvtot{}{x} \left[\frac{1}{L^0(x)} \dvpart{\psi(x,t)}{x}\right] = C^0(x) \ddot \psi(x,t)\\
\ee
is solved by first decomposing $\psi(x,t)$ over (unitless) normal modes $u_n(x)$,
\be
\label{eqn:decomp-spec}
\psi(x,t) = \sum_n \psi_n(t) u_n(x).
\ee
Here $\psi_n$ is the flux amplitude of eigenmode $n$, of frequency $\omega_n$ and eigenfunction $u_n(x)$ and is given by
\be
\psi_n(t) = \frac{1}{N} \int_{-\ell}^{+\ell} C^0(x)  \psi(x,t) u_n(x) dx,
\ee
where $N$ is a normalization constant. Assuming a large quality factor $Q$, the current at the two ends of the resonator vanish; the eigenmodes must satisfy the boundary conditions $\partial_x u_n(x = \pm \ell) = 0$. Spectral decomposition of the flux $\psi(x,t)$ in Eq.~\eqref{eqn:euler-lagrange} leads to a Sturm-Liouville differential equation of the form
\be
\label{eqn:sturm-liouville}
\dvtot{}{x} \left[\frac{1}{L^0(x)} \dvpart{u_n(x)}{x}\right] = -\omega_n^2 C^0(x) u_n(x),
\ee
whose solutions $\{u_n(x),\omega_n\}$ form an orthogonal basis. The eigenfunctions $u_n(x)$ respect a weighted orthogonality relation
\be
\label{eqn:ortho-chi}
\int_{-\ell}^{+\ell} C^0(x) u_n(x) u_m(x) dx = C_r \delta_{nm},
\ee
where the normalization constant is chosen to be the total capacitance of the transmission-line, $C_r = \int_{-\ell}^{\ell} C^0(x) dx$. 

\begin{figure}[t]
\centering
\includegraphics[width=0.35\textwidth]{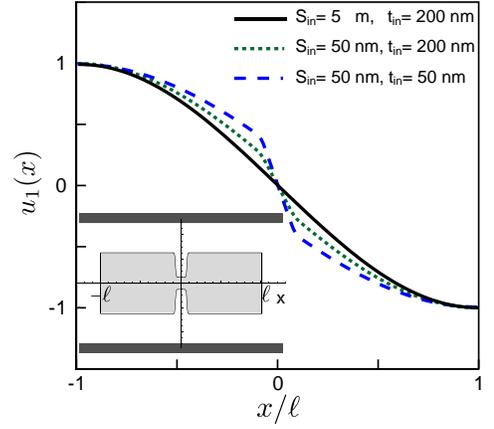}
\caption[]{(Color online) First mode (normalized to 1) for aluminum transmission-line resonators with the different constriction parameters listed in Table~\ref{table:g}. As the central line is reduced in cross section, the slope of the flux field increases inside the constriction. The inset shows the geometry of the constriction in the center conductor of the resonator.}
\label{fig:data-case-all}
\end{figure}

\begin{table*}
\caption{Inductance ratio $L^0_\mathrm{in}/L^0_\mathrm{out}$ inside and outside the constriction, resonant frequency $\omega_1/2\pi$ of the first mode, flux gradient inside the constriction $\left|\partial \psi_1(x)/\partial x\right|_{x=0}$ and qubit-resonator flux coupling $g_{\hat \varphi,1}^{ge}$ at $\Phi_{ext} = \Phi_0/2$ for different values of the resonator center conductor width $S_\mathrm{in}$ and thickness $t_\mathrm{in}$ at the location of the constriction assuming a length of $w = 5 ~\mu$m of the shared part between the qubit and resonator. Qubit parameters are given in the text while resonator parameters are given in Appendix~\ref{sec:inhomogeneousTL}.}
\begin{ruledtabular}
\begin{tabular}{c c c c c c c}
& $S_\mathrm{in}$ & $t_\mathrm{in}$ &  & $\omega_1/2\pi$ & $ \left|\partial \psi_1(x)/ \partial x\right|_{x=0}$  & $g_{\hat\varphi,1}^{ge}/2\pi $ \\

Material & (nm) & (nm) & $L^0_\mathrm{in}/L^0_\mathrm{out}$ & (GHz) & $(10^{-6} \Phi_0/\mu$m) & (MHz) \\

\hline
Al& 5000 & 200 & 1 & 13.12 & 12.97 & 71.8 \\

Al & 50 & 200 & 3.4 & 10.98 & 35.36 & 195.6 \\

Al & 50 & 50 & 4.1 & 10.52 &  40.22 & 222.5\\ 

Nb & 50 & 50 & 8.3 & 8.62 &  65.10 & 360.2\\ 
\end{tabular} 
\end{ruledtabular}
\label{table:g}
\end{table*}

By using the spectral decomposition~\eqref{eqn:decomp-spec} in the Lagrangian density~\eqref{eqn:lag-density} and using the orthogonality relations~\eqref{eqn:ortho-chi} along with the Sturm-Liouville differential equation~\eqref{eqn:sturm-liouville}, the total Lagrangian simplifies to a sum over eigenmodes:
\be
\mathcal L =  \sum_{n} \frac{C_r}{2} \dot \psi_n^2 - \frac{C_r}{2} \omega_n^2 \psi_n^2.
\ee
Defining the charge $\theta_n= C_r \dot \psi_n$ as the conjugate momentum to the flux $\psi_n$, the corresponding Hamiltonian is
\be
\mathcal H= \sum_n \frac{\theta_n^2}{2C_r} + \frac{C_r}{2} \omega_n^2 \psi_n^2.
\ee
By quantifying and introducing the operators
\be
\label{eqn:cavity_operators}
\bsplit
\hat \psi_n =& \sqrt{\frac{\hbar}{2 \omega_n C_r}} (a_n^\dagger+ a_n),\\
\hat \theta_n =&i \sqrt{\frac{\hbar \omega_n C_r}{2}} (a_n^\dagger- a_n),\\
\end{split}
\ee
with $[a_n,a^\dagger_m]=\delta_{nm}$, we arrive at the standard form
\be
\label{eq:Hoscillators}
\mathcal H_\mathrm{tl} = \sum_n \hbar \omega_n \left(a_n^\dagger a_n + 1/2 \right),
\ee
completing the mapping of the inhomogeneous resonator to a sum of harmonic oscillators. Unlike the homogeneous case, the mode frequencies can be \emph{inharmonically} distributed such that the equality $\omega_n = n \omega_0$ is not satisfied in general. Like most Sturm-Liouville problems, the eigenmodes $u_n(x)$ and eigenfrequencies $\omega_n$ are found numerically by exact diagonalization~\cite{ledoux:2005a}. As discussed in Appendix~\ref{sec:inhomogeneousTL}, details of the transmission-line geometry are important in determining these quantities. 

Figure~\ref{fig:data-case-all} shows the first mode $u_1(x)$ for three different configurations of a constriction in the center conductor of the resonator, as detailed in Table~\ref{table:g}.  As the constriction is made narrower, and thus the local inductance made larger, an abrupt change in $u_1(x)$ develops.  A flux qubit connected on either side of the constriction, as illustrated in Fig.~\ref{fig:qubit}, will thus be strongly phase biased.  As a result, an inhomogeneity in the resonator can increase the qubit-resonator coupling, making the strong coupling regime easier to reach. 

Qubit-resonator coupling can also benefit from the kinetic inductance of superconducting materials with large London penetration depth $\lambda_L$, such as niobium where $\lambda_L = 39$~nm, when the dimensions of the cross-section of the central line reaches dimensions comparable with $\lambda_L$. Alternatively, a Josephson junction with large Josephson inductance $L_J \sim \varphi_0^2/E_J$ can replace the constriction shown in Fig.~\ref{fig:cavite-qubit} to provide even stronger coupling. In Appendix ~\ref{sec:TLplusJJ}, we show how in this case, the field $\psi(x,t)$ becomes discontinuous at the location of the junction and presents an important flux difference across the junction. We also show how the Hamiltonian of the transmission-line plus junction can also be written in the standard form of Eq.~\eqref{eq:Hoscillators}~\cite{bourassa:2009a}.  Consequences of this very large coupling will be discussed further below.

\section{Qubit-resonator Hamiltonian}

In this section, we obtain the qubit-resonator Hamiltonian for the system of Fig.~\ref{fig:cavite-qubit} focusing on the case of the inhomogeneous resonator.  Derivation of the Hamiltonian in the presence of a fourth junction instead of a constriction can be done following these lines and is discussed in Appendix~\ref{sec:TLplusJJ}.  Figure~\ref{fig:qubit} shows in more details the qubit connected to the center conductor of the resonator.  Including the flux qubit, the Lagrangian reads
\be\label{eq:Lagrangian:qubit_resonator}
\bsplit
\mathcal L & 
= \mathcal L_\mathrm{tl} 
+ \sum_{k=1}^3 \left\{ \frac{C_{Jk}}{2}\dot\phi^2_k + E_{Jk} \cos [\phi_k/\varphi_0] \right\},
\end{split}
\ee
where $C_{Jk}$ is the capacitance of junction $k$,  $E_{Jk}$ its Josephson energy and $\phi_k$ the flux difference across it, and $\varphi_0=\Phi_0/2\pi$ is the reduced flux quantum. Junctions 1 and 3 are assumed to be equivalent, $C_{J1} = C_{J3} \equiv C_J$ and $E_{J1} = E_{J3} \equiv E_J$, while junction 2 is such that $C_{J2} = \alpha C_J$ and $E_{J2}= \alpha E_J $ with $\alpha < 1$~\cite{orlando:1999a}. The flux differences $\phi_{1(2)}$ depend explicitly on the resonators voltage through $\psi(x_{1(2)})$, with $x_{1}$ and $x_{2}$ the positions of the resonator where the qubit loop is connected. Finally, the phase differences satisfy
\be\label{eqn:constraint}
\phi_1 + \phi_2 - \phi_3 - \psi_\mathrm{tl} = \Phi_\mathrm{ext},
\ee
where $\psi_\mathrm{tl}  = \psi(x_2) - \psi(x_1)$ and $\Phi_\mathrm{ext}$ is an externally applied flux. This constraint is used to eliminate $\phi_2$ from Eq.~\eq{eq:Lagrangian:qubit_resonator}.

\begin{figure}[t]
\includegraphics[width=0.40\textwidth]{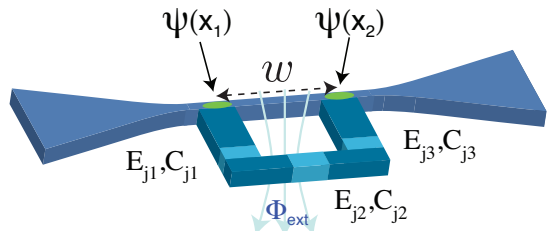}
\caption[]{(Color online) Closeup of the flux qubit fabricated at the location of the constriction.  The qubit is attached at positions $x_1$ and $x_2$ on the resonator.  $\Phi_\mathrm{ext}$ is an externally applied flux. A large Josephson junction inserted in the center conductor of the resonator between $x_1$ and $x_2$ can lead to stronger coupling.}
\label{fig:qubit}
\end{figure}

In obtaining the Hamiltonian, we assume that the qubit does not significantly perturb the resonator such that the mode decomposition for $\psi(x,t)$ found in the previous section is a good approximation even in the presence of the qubit.  This approximation is accurate for small qubit capacitances such that the capacitive terms in Eq.~\eqref{eq:Lagrangian:qubit_resonator} do not induce large frequency shifts of the resonator, and if the inductance of the center line of the resonator of length $w=x_{2}-x_{1}$ where the qubit is connected is smaller than the total inductance of the qubit [i.e.~$L^0(x_1)w/\sum_k L_{j,k} \ll 1$ with $L_{j,k}=\varphi_0^2/E_{Jk}$ the Josephson inductance of junction $k$) such that most of the current is flowing through the resonator.  Both of these assumptions can safely be satisfied in practice with small junctions.  We note that while these constraints are useful in deriving the system Hamiltonian, the main results can hold even if they are not strictly respected.

It is useful to introduce the sum and difference fluxes $\phi_\pm = \{[\phi_3 + \psi(x_2)]\pm[\phi_1+\psi(x_1)]\}/2$, where $\phi_1 + \psi(x_1)$ and $\phi_3 + \psi(x_2)$ represent the flux on the island of the qubit separated from the resonator by junction 1 and 3, respectively. The charges conjugate to these fluxes are $q_\pm = \partial \mathcal L/ \partial \dot \phi_\pm$.  Using these conjugate variables, the Hamiltonian is easily obtained in the usual way~\cite{devoret:1997a}. After transformation under the unitary $T_+ T_-$, with
\be
\label{eqn:unitary}
T_\pm =\prod_n \text{exp}\left[\frac{-i}{2\hbar}  \psi_n q_\pm \delta_n^\pm \right],
\ee
where $\delta_n^{\pm} = u_n(x_2) \pm u_n(x_1)$ and using phase variables $\varphi_\mu = \phi_\mu/\varphi_0$, the Hamiltonian reads
\be
\label{eqn:ham-trans}
\bsplit
H &
= 
\sum_n 
\left[
\hbar \omega_n a^\dagger_n a_n 
+ \frac{ q_-^2}{2C_n^-} + \frac{ q_+^2}{2C_n^+} 
-\frac{2C_{j2}}{\wt C_n^2} \delta_n^- q_- \hat \theta_n 
\right]\\
& 
-E_{J}
\left[
2  \cos \varphi_+ \cos \varphi_- 
+ \alpha \cos( \varphi_{ext}  + \hat \varphi +  2 \varphi_- )
\right].
\end{split}
\ee
In this expression, $\hat \varphi$ is a quantum flux bias given by
\be
\hat \varphi = \sum_n \hat \psi_n \delta_n^-/\varphi_0,
\ee
where in general $|\hat \varphi|\ll 1$, and the resonator mode-dependent capacitances are
\be
\bsplit
\frac{1}{C_n^-} =&  \frac{2 C_r +2 C_{j2} (\delta_n^-)^2}{\wt C_n^2}\\
\frac{1}{C_n^+}=&  \frac{C_r (C_{j1}+2C_{j2}) + C_{j1}C_{j2} (\delta_n^-)^2}{\wt C_n^2 C_{j1}}\\\end{split}
\ee
with $\wt C_n^2 = 2[C_r( C_{j1} + 2C_{j2} ) + (\delta_n^-)^2C_{j1}C_{j2}]$. The Josephson potential energy takes the usual form for a 3-junction flux qubit~\cite{orlando:1999a}. The usefulness of the unitary transformation is to change the phase bias from vacuum fluctuations in the resonator field to a flux bias $\hat \varphi$ directly on the qubit which is simply adding to the external flux $\varphi_\mathrm{ext} = \Phi_{\mathrm{ext}}/\varphi_0$. 

Defining the qubit capacitances $C^\pm =\sum_n  C_n^\pm$ and expanding the term proportional to $\alpha E_J$ to first order in $\hat \varphi$, the resulting Hamiltonian becomes
\be
H = H_r + H_{qb} + H_{\hat q} + H_{\hat\varphi},
\ee
where $H_r = \sum_n \hbar\omega_n a^\dag_n a_n$ is the resonator Hamiltonian, 
\be
\begin{split}
H_{qb} & =  
\frac{q_-^2}{2C^-} + \frac{q_+^2}{2C^+} \\
&
-E_{J} \left[ 2  \cos \varphi_+ \cos \varphi_- + \alpha \cos( \varphi_{ext}  +  2  \varphi_- ) \right]
\end{split}
\ee
the standard flux qubit Hamiltonian~\cite{orlando:1999a}, and
\begin{align}
H_{\hat q} &= -\sum_n \frac{2C_{j2}}{\wt C_n^2} \delta_n^- q_- \hat \theta_n, \\
H_{\hat\varphi} &=   \alpha E_{J} \hat \varphi  \sin (\varphi_{ext} +2 \varphi_-)
\end{align}
describe charge and flux coupling of the qubit to mode $n$ of the resonator, respectively.

Projecting on the eigenstates $\{\ket k\}$ of frequencies $\{\Omega_k\}$ of the qubit Hamiltonian $H_{qb}$, the flux coupling Hamiltonian $H_{\hat\varphi}$ can be expressed as
\be 
\label{eq:H-non-RWA}
H_{\hat \varphi} = \sum_n \sum_{k,l}\hbar g_{\hat \varphi,n}^{kl} \ket k \bra l \left(a^\dagger_n+a_n\right)
\ee
where
\be
\label{eqn:flux-coupling}
\bsplit
\hbar g_{\hat \varphi,n}^{kl} = \alpha E_J \Delta \varphi_n \bra k \sin(\varphi_{ext} + 2 \hat \varphi_-)\ket l.
\end{split}
\ee
Here we have used Eq.~\eqref{eqn:cavity_operators} and defined $\Delta \varphi_n = \delta_n^- \sqrt{\hbar/2C_r \omega_n}/\varphi_0$. These matrix elements are easily evaluated after diagonalizing $H_{qb}$ numerically to find the exact qubit eigenstates.  At the flux sweet-spot, $\Phi_{\mathrm{ext}} = \Phi_0/2$, only off-diagonal coupling $g_{\hat \varphi,n}^{kl}$ between states $k$ and $l$ of different parity remain~\cite{liu:2005a}.  

Using the expression of Eq.~\eq{eqn:cavity_operators} for $\hat\theta_n$, the above selection rule reduces the charge coupling $H_{\hat q}$ to
\be
H_{\hat q} = \sum_n \sum_{k,l > k } \hbar g_{\hat q,n}^{kl}  (\ket k \bra l - \ket l \bra k) \left(a^\dagger_n - a_n\right),
\ee
for states $\ket k, \ket l$ of different parity and where
\be
\hbar g_{\hat q,n}^{kl} =  \frac{2C_{j2} \delta_n^-}{i\wt C_n^2} \sqrt{ \frac{\hbar  C_r \omega_n}{2} }  \bra{k}  q_- \ket{l}
\ee
is a real quantity and maximal between states $(k,l)=(1,2)$.   Comparing $g_{\hat q}$ to $g_{\hat \varphi}$ we get
\be
\left| \frac{g^{kl}_{\hat \varphi,n}}{g^{kl}_{\hat q,n}}\right|
 \approx 
 \frac{ (2 \alpha+1)  E_{J} } {\hbar \omega_n} 
 \frac{\bra k\sin(\varphi_{ext} + 2 \varphi_-)\ket l}{ \bra k q_-/2e \ket l }.
\ee
Since in practice $E_{j} \gg \hbar \omega_n$ for flux qubit, the charge matrix elements are at best a fraction of unity in the vicinity of flux degeneracy point, we find $\left|g^{k,l}_{\hat \varphi}/g^{k,l}_{\hat q}\right| \geq 10^2-10^3$.  Unsurprisingly, charge coupling is negligible.

\section{Jaynes-Cummings Hamiltonian}

\begin{figure}[t]
\includegraphics[width=0.4\textwidth]{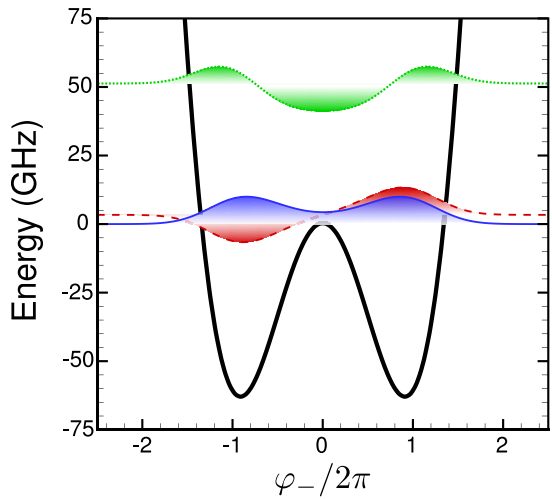}
\caption[]{(Color online) Cut along the $\varphi_+=0$ axis of the double well potential of the flux qubit with the first (full blue), second (dashed red) and third (dotted green) eigenstates for qubit parameters $E_{J1}=259$~GHz, $\alpha=0.8$ and $E_J/E_C =35$ with $\Phi_{\mathrm{ext}} = \Phi_0/2$. }
\label{fig:qubit_potential}
\end{figure}

In the rotating-wave approximation (valid when $g^{kl}_{\hat \varphi,n} \ll \{ \omega_n,\omega_{\mathrm{min}(k,l)}\}$), the full Hamiltonian takes the Jaynes-Cummings form~\cite{walls:1994a}
\be
\label{eq:H-RWA}
H = \sum_n \hbar\omega_n a_n^\dag a_n + \sum_k \hbar \Omega_k \ket k \bra k + \sum_{n,k,l} \hbar g_{n}^{kl} (\ket k \bra l a^\dag_n + \mathrm{h.c.}),
\ee
where $g_n^{kl}=g_{\hat\varphi,n}^{kl} -  g_{\hat q,n}^{kl} \approx g_{\hat\varphi,n}^{kl}$.

In addition to the strong coupling and the low charge noise, an important advantage of studying the Jaynes-Cummings physics in this system is the very large anharmonicity of the flux qubit compared to the transmon~\cite{koch:2007a}.  This is illustrated in Fig.~\ref{fig:qubit_potential} which shows the first few eigenenergies of the flux qubit Hamiltonian $H_q$.  Moreover, with flux qubits it is also possible to take advantage of the fact that the first two eigenstates can be localized in the wells of the potential, while the higher eigenstate is delocalized.  These three states can be used as a $\Delta$ system~\cite{liu:2005a}, opening possibilities for many quantum optics phenomena in cavity QED with superconducting circuits.

It is also worth pointing out that multiple qubits can be coupled to the same resonator.  With the qubits fabricated in close proximity to a node of the eigenfunction to which they are (most strongly) coupled, the Hamiltonian of the system simply reduces to
\be
H = H_r + \sum_{j=1}^N \left[ H_{qb}^j + H_{\hat q}^j + H_{\hat \varphi}^j\right]
\ee
where the coupling Hamiltonians are calculated by projecting the operators onto each qubit subspace.  Two-qubit gates can be generated in this system in the same way as with charge-qubit based circuit QED~\cite{blais:2007a}.

\section{Comparison to geometric coupling and numerical results}

As shown in the last section, qubit-resonator flux coupling is provided by the vacuum fluctuations of the resonator field  $\hat \varphi$ threading the qubit loop.  For a qubit sharing a length $w$ with the resonator, this flux can be expressed as
\be
\label{eqn:flux}
\begin{split}
\hat \varphi = 
& \sum_n 
 \sqrt{\frac{\hbar}{2 \omega_n C_r}}  \left[u_n(x_1+w) - u_n(x_1)\right]
 (a^\dagger_n +a_n) /\varphi_0\\
& \approx - w L_0(x_1) \hat I(x_1)/\varphi_0,
\end{split}
\ee
where $L_0(x_1)$ and $\hat I(x_1)  = -\partial_x \hat \psi(x)/L_0(x)|_{x_1}$ are, respectively, the inductance per unit length and the current in the resonator at the location $x_1$ of the qubit.  The slope of the flux field $\partial_x \hat \psi(x)$ at the location of the qubit is given in Table~\ref{table:g} for different resonator geometries and materials.  As can be seen there, coupling is enhanced by locally increasing the resonator inductance. This is to be compared to the case where the qubit loop is mutually coupled to the resonator only by mutual inductance~\cite{yang:2005a,lindstrom:2007a}. In this situation, one finds $\hat \varphi = M \hat I(x_1)/\varphi_0$ with
\be
M \sim \frac{\mu_0 w}{2\pi} \ln\left[ \frac{d+L}{d}\right]
\ee
as given from the Neumann formulae for a rectangular loop of length $w$ and width $L$ separated by a distance $d$ from the resonator center conductor, here approximated by a infinite cylindrical wire. Geometric coupling will win over direct coupling only if the ratio $L/d$ is made such that
\be
 \frac{L}{d}  \gtrsim e^\frac{L^0(x_1)}{\mu_0/2\pi} - 1.
\ee
Since $L^0(x_1)$ can reach several units of $\mu_0/2\pi$ in the constriction due to the contribution of the kinetic inductance, geometric coupling can only win by either increasing the qubit area $(w \times L)$ or by reducing the distance between the qubit loop and the resonator central line $(d)$. Large loops will make the qubit more susceptible to surrounding flux noise, while placing the qubit very close to the resonator can be challenging in addition to increasing the capacitive coupling to the resonator.  In contrast, direct coupling leads to a coupling strength that scales with length $w$ of the shared part between qubit and resonator rather than with the area.  Strong coupling can therefore be reached without large sensitivity to flux noise.

Table~\ref{table:g} shows the coupling $g_{\hat\varphi,n}^{ge}$ of directly connected qubits for various transmission-line configurations.  To obtain these results, we have taken parameters close to those of Ref.~\cite{chiorescu:2003a} with $E_{J1} = 259$ GHz, $\alpha = 0.8$ and $E_J/E_C=35$.  These parameters where used in Fig.~\ref{fig:qubit_potential}. The qubit's loop width was taken to be $w = 5\:\mu$m.  For these realistic values, this system easily reaches the strong coupling regime. Moreover, in all cases shown here, the coupling strength is substantially larger than the estimates for geometric coupling~\cite{yang:2005a,lindstrom:2007a}. 

\section{Ultra-Strong Coupling regime}

\begin{figure}
\centering
\includegraphics[width=0.40\textwidth]{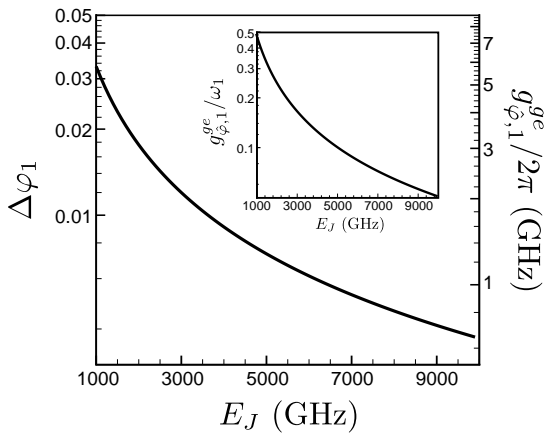}
\caption[]{Flux bias $\Delta \varphi_1$ and resultant qubit-resonator coupling energy $g^{ge}_{\hat \varphi,1}$ as calculated from Eq.~\eqref{eqn:flux-coupling} induced by a Josephson junction placed in the center of a homogenous aluminum transmission-line resonator (see Table~\ref{table:g}) as a function of the Josephson energy $E_J$. The inset shows that the qubit-resonator coupling energy can reach several tens of percent of the resonator frequency. Qubit parameters are given in the text.}
\label{fig:flux-diff-jj}
\end{figure}

Using a Josephson junction to locally change the inductance can result in much stronger coupling.  As illustrated in Fig.~\ref{fig:flux-diff-jj}, for a relatively large Josephson energies $E_J \geq 1000$~GHz, the phase bias seen by the qubit is so large that the coupling energy can easily reach $g^{ge}_{\hat \varphi,1}/2\pi \sim 1000$~MHz and beyond, corresponding to several tens of percent of the resonator frequency.   This coupling can be increased further by lowering the Josephson energy of the inserted junction, as long as the corresponding Josephson inductance is small compare to that of the qubit.  

In this ultra-strong coupling regime~\cite{ciuti:2005a,ciuti:2006a}, the RWA, used in going from Eq.~\eq{eq:H-non-RWA} to Eq.~\eq{eq:H-RWA}, breaks down and the full Hamiltonian must be considered.  This circuit then becomes a solid-state implementation of the $E \otimes \beta$ Jahn-Teller model~\cite{jahn:1937a}. While coupling of the artificial atom to the electric field of the resonator does not lead to super-radiant phase transition, magnetic coupling, which dominates here, does~\cite{knight:1978a,larson:2008a}. In this situation, the ground state of the combined qubit-resonator system can be an entangled state corresponding to the oscillator being displaced by a qubit-state dependent quantity~\cite{hines:2004a,meaney:2009a}. In this ground state, we can expect a finite photon population, something which could be measured using a second resonator in a number splitting experiment~\cite{gambetta:2006a,schuster:2007a}.

In addition to the breakdown of the RWA, higher-order terms in the expansion of the Josephson energy in Eq.~\eq{eqn:ham-trans} have to be taken into account as the coupling strength increases.  Second-order corrections lead to an additional flux coupling Hamiltonian $H_{\hat \varphi}^{(2)}$ of the form
\be
H_{\hat \varphi}^{(2)} = \sum_n \hbar \zeta_{\hat \varphi,n}^{kl} \ket k \bra l \left(a^\dagger_n +a_n\right)^2
\ee
where
\be
\hbar \zeta_{\hat \varphi,n}^{kl} = \frac{\alpha}{2} E_J (\Delta\varphi_n)^2 \bra k \cos(\varphi_{ext} + 2\hat\varphi_-)\ket l
\ee
This second-order correction leads to ac-Stark shifts and, more interestingly, can be used to generate squeezing of the microwave field inside the resonator. Tuning the circuit parameters can lead to detectable effects with $\zeta_{\hat \varphi,n}^{kl}/2\pi \gtrsim 1$ MHz. We note that this also leads to resonator mode-mode coupling.  In practice however, the frequency separation between these modes is large enough that this can be neglected.

\section{Conclusion}

We have obtained the Hamiltonian of a superconducting flux-qubit directly coupled to the center conductor of a coplanar transmission-line resonator.  By using a constriction in the center line of the resonator, the coupling strength between the qubit and the resonator can be significantly increased.  This is due to the increase in the geometric and kinetic inductance of the line and the resulting large phase bias seen by the qubit.  There are two main advantages of this approach compared to coupling based on the mutual inductance between the qubit and the resonator: the coupling is much stronger in magnitude and this is possible without working with large qubit loops which would increase sensitivity to flux noise.  Together with the insensitivity of flux qubits to charge noise, its large anharmonicity and its $\Delta$ configuration, this approach leads to the possibility of studying numerous quantum optics effects with superconducting circuits. Finally, by replacing the constriction with a Josephson junction of large Josephson inductance, we have shown that the coupling can be as large as several tens of percent of the resonator frequency.  In this situation, the breakdown of the RWA should lead to an entangled qubit-resonator ground state.

\begin{acknowledgments}
J.B. was supported by NSERC and FQRNT. 
A.B. was supported by CIFAR, NSERC  and the Alfred P. Sloan Foundation.
J.M.G. was supported by NSERC, CIFAR, MRI and MITACS.
A.A.A., O.A. and Y.N. were supported by CREST program of the Japan Science and Technology Agency (JST).

\end{acknowledgments}

\appendix

\section{Design and characteristics of inhomogeneous transmission lines}
\label{sec:inhomogeneousTL}

Appropriate modeling of the inhomogeneous transmission-line electrical characteristics is needed to compute  eigenmodes, frequencies and ultimately the  coupling between the qubit and the resonator. In this section, we give details on the geometry of the inhomogeneous resonators that were used in Table~\ref{table:g}.

\begin{figure}[t]
\centering
\includegraphics[width=0.3\textwidth]{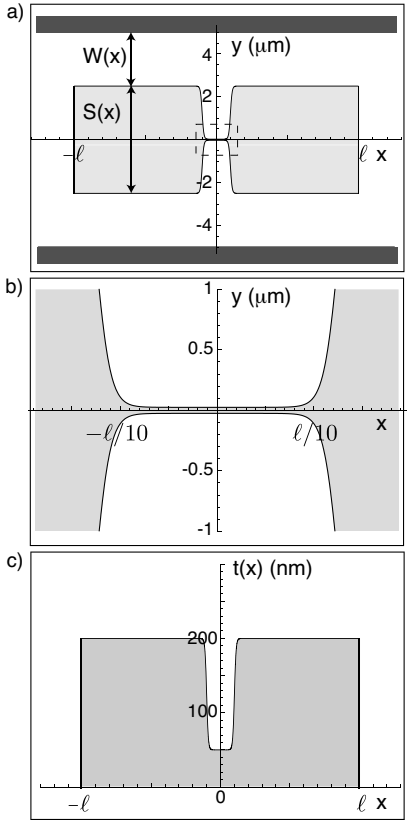}
\caption[]{Geometry of the inhomogeneous transmission line used in numerical simulations. Central-line width $S(x)$ and ground-plane spacing $W(x)$ are shown in a) for the total length of the line $2\ell = 5$ mm, and in b) around the constriction indicated by the dashed square in a). The thickness of the central-line is shown in c). In this example, the outer dimensions of the homogeneous part of the line are $S_\mathrm{out} = 5$ $\mu$m, $W_\mathrm{out} = 2.5$ $\mu$m and $t_\mathrm{out} = 200$ nm while the inner dimensions inside the constriction at $x=0$ with $S_\mathrm{in} = 50$ nm, $W_\mathrm{in}=4.95$ $\mu$m and $t_\mathrm{in} = 50$ nm.}
\label{fig:an-const-final}
\end{figure}

The capacitance per unit length $C^0(x)$, the inductance per unit length $L^0(x)$ and the impedance $Z^0(x)$ of the coplanar transmission-line resonator depend on the ratio between the width of the center electrode $S$ and the distance between the two ground planes $S + 2W$, $W$ being the distance between the ground plane and the edge of the central line~\cite{simons:2001a}:
\be
\bsplit
C^0 =& 2 \epsilon_0(\epsilon_r +1) \frac{K(k_0)}{K(k'_0)}, \hspace{0.5cm} L^0_{\mathrm{geo}} = \frac{\mu_0}{4} \frac{K(k'_0)}{K(k_0)},\\
Z^0=& \sqrt{\frac{L^0_{\mathrm{geo}} + L^0_{\mathrm{kin}}}{C^0}},
\end{split}
\ee
where $\epsilon_r$ is the dielectric constant of the subtrate, $k_0 = S/(S+2W)$ is the aspect ratio, $k'_0 = \sqrt{1-k_0^2}$ and $K(x)$ is the complete elliptic integral of the first kind. By decreasing the aspect ratio $k_0$ along the line, the inductance and impedance of the line are locally increased while the capacitance is decreased. For a superconducting resonators, the kinetic inductance  can be expressed as \cite{watanabe:1994a}
 \be
 \bsplit
 L^0_{\mathrm{kin}} =& \mu_0 \lambda_L(T) \frac{C}{4 ADK(k_0)} \left[ \frac{1.7}{\sinh(t/2\lambda_L(T))} \right.\\
 & \left.+ \frac{0.4}{\sqrt{[(B/A)^2 -1][1-(B/D)^2]}}\right]
 \end{split}
 \ee
 where $\lambda_L(T)$ is the London penetration depth of the superconductor at temperature $T$, $t$ is the thickness, and 
 \be
 \bsplit
 A =& -\frac{t}{\pi} + \frac{1}{2} \sqrt{\left(\frac{2t}{\pi}\right)^2 + S^2},\hspace{0.5cm} B = \frac{S^2}{4A},\\
 C=& B - \frac{t}{\pi} +\sqrt{\left(\frac{t}{\pi}\right)^2 + W^2}, \hspace{0.4cm}D= \frac{2t}{\pi} + C.\\
 \end{split}
 \ee
As for the geometrical inductance, a decrease of the aspect ratio will increase the kinetic inductance of the line but the effect is rather marginal unless the dimensions of the cross section of the central electrode become of the order of $\lambda_L$.

The geometry of the inhomogeneous transmission-line used in Table~\ref{table:g} is depicted in Fig.~(\ref{fig:an-const-final}). We consider a regular, initially homogeneous transmission line resonator made of either aluminum ($\lambda_L =16$ nm)  or niobium ($\lambda_L = 39$ nm) with a total length of $2\ell = 5$ mm. The central electrode is $t=200$ nm thick and $S=5$ $\mu$m wide. The ground planes are $W=2.5$ $\mu$m away from the edge of the central-line. For simplicity, the distance between the ground planes ($S+2W$) is not modified. The width and thickness of the central electrode are reduced at the center of the resonator to create a constriction. The dimensions are reduced gradually over a length $d$ from the initial \textit{outer} values $S_\mathrm{out}$, $W_\mathrm{out}$ and $t_\mathrm{out}$ down to the minimal \textit{inner} values $S_\mathrm{in}$, $W_\mathrm{in}$ and $t_\mathrm{in}$ at $x=0$. The electrode width $S(x)$, thickness $t(x)$ and ground-planes spacing $W(x)$ are continuous smooth functions for ease of computation.

Table~\ref{table:g} summarizes the numerical results obtained for inhomogeneous resonators made of Al and Nb and characterized by width, thickness and ground-plane spacing as illustrated in Fig.~\ref{fig:an-const-final}. For aluminum, as the central line cross section dimensions are reduced from $5$~$\mu$m $\times$ $200$ nm down to $50 \times 50$ nm$^2$, the total inductance per unit length can be increased by a factor of 4. As it shown in Fig.~\ref{fig:data-case-all}, the slope of the flux field $|\partial_x \psi_1(x)| = |\partial_x u_1(x)|\sqrt{\hbar/2C_r \omega_1}$ inside the constriction is also increased by a factor of 4. On the other hand, because of the larger London penetration depth, the kinetic inductance of niobium resonators can be very important.  In this case,  the local inductance can be increased by a factor of $\sim 8$ for a cross-section of 50 $\times$ 50 nm$^2$, leading to a fivefold increase in the slope of the flux field.

\section{Transmission line intersected by a Josephson Junction}
\label{sec:TLplusJJ}

\begin{figure}[t]
\centering
\includegraphics[width=0.35\textwidth]{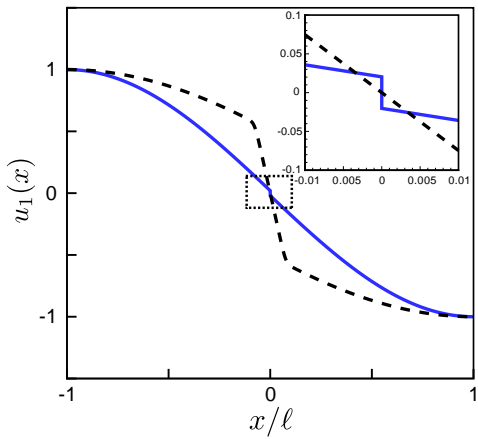}
\caption[]{(Color online) First mode (normalized to 1) for the niobium transmission line (dashed black) and for a homogeneous aluminum transmission-line resonator intersected by a (pointlike) Josephson junction of plasma frequency $\omega_p/2\pi = 40$~GHz with $E_J = 6000$ GHz (full blue). The transmission-line characteristics are detailed in Table~\ref{table:g}. The inset shows an enlarged view of the mode in the vicinity of the junction. The abrupt variation in the mode at the position of the junction enables a much greater flux bias on the qubit.}
\label{fig:mode-const-jj}
\end{figure}

Inserting a Josephson junction in the center conductor of the resonator at the location of the qubit can lead to significantly stronger coupling.  We note that having a four, rather than three, junctions is natural for a flux qubit~\cite{burkard:2005c}.  If the Josephson inductance of the resonator junction is much smaller than the total inductance of the qubit loop, the qubit acts once again as a simple perturbation on the resonator eigenmodes. The theoretical description of a resonator with an integrated Josephson junction can be found elsewhere~\cite{bourassa:2009a} and we recall only the main results (see Fig.~\ref{fig:mode-const-jj}).

For a transmission-line resonator interrupted by a Josephson junction of linear Josephson inductance $L_J = \varphi_0^2/E_J$ and capacitance $C_J$, the eigenmodes can be described by spatially oscillating functions $u_m(x)$ given by
\be
u_m(x) = A_m \begin{cases}
					 \cos[k_m(x+\ell )] & x < x_j \\
				B_m \cos[k_m (x-\ell )] & x > x_j .
				  \end{cases}
\ee
If the junction is placed at the center $x=0$ of the resonator, $B_m=-1$ and the wave-vectors $k_m$ are solutions of the transcendental eigenvalue equation 
\be
\frac{2L_0}{L_J} \left(1 - \frac{\omega_m^2}{\omega_p^2}\right) \cot[k_m \ell ] =  k_m,
\ee
where $\omega_m = k_m/\sqrt{L^0 C^0}$ are the resonance frequencies of the circuit and $\omega_p = 1/\sqrt{L_J C_J}$ is the plasma frequency of the junction. The eigenmodes of the circuit are found to obey a generalized orthogonality equation 
\be
 \int_{-\ell}^{\ell}  dx C_0 u_m(x) u_{m'}(x)  + C_J \delta^-_m \delta^-_{m'}  = C_\Sigma \delta_{mm'},
\ee
where $\delta^-_m = u_m(0^+) - u_m(0^-)$ is the dimensionless mode gap at the junction and $C_\Sigma = 2 \ell C^0 + C_J$ is the total capacitance of the circuit. This orthogonality equation is used to fix the normalization $A_m$. 

It follows that in the linear approximation of the Josephson inductance, the resonator can be describe by sum of harmonic oscillators of frequency $\omega_m$
\be
H = \sum_m \hbar \omega_m \left(a^\dagger_m a_m  + 1/2\right)
\ee
where the ladder operators obey the commutation relation $[ a_m ,a^\dagger_{m'}] = \delta_{mm'}$ and define flux $\hat \psi_m$  and charge $\hat q_m$ operators given by Eq.~\eqref{eqn:cavity_operators} with the appropriate capacitance and frequency definitions. As it is shown in Fig.~\ref{fig:mode-const-jj}, the presence of the junction creates a very abrupt discontinuity in the modes.  This leads to a large flux field slope $\partial_x \hat \psi_1(x)$, which in turns leads to very strong qubit-resonator coupling.  We note that very strong coupling can be attained even for negligible non-linearity of the resonator mode~\cite{bourassa:2009a}.

\singlespacing

\begin{thebibliography}{21}
\expandafter\ifx\csname natexlab\endcsname\relax\def\natexlab#1{#1}\fi
\expandafter\ifx\csname bibnamefont\endcsname\relax
  \def\bibnamefont#1{#1}\fi
\expandafter\ifx\csname bibfnamefont\endcsname\relax
  \def\bibfnamefont#1{#1}\fi
\expandafter\ifx\csname citenamefont\endcsname\relax
  \def\citenamefont#1{#1}\fi
\expandafter\ifx\csname url\endcsname\relax
  \def\url#1{\texttt{#1}}\fi
\expandafter\ifx\csname urlprefix\endcsname\relax\def\urlprefix{URL }\fi
\providecommand{\bibinfo}[2]{#2}
\providecommand{\eprint}[2][]{\url{#2}}

\bibitem[{\citenamefont{Schoelkopf and Girvin}(2008)}]{schoelkopf:2008a}
\bibinfo{author}{\bibfnamefont{R.~J.} \bibnamefont{Schoelkopf}}
  \bibnamefont{and} \bibinfo{author}{\bibfnamefont{S.~M.}
  \bibnamefont{Girvin}}, \bibinfo{journal}{Nature (London)}
  \textbf{\bibinfo{volume}{451}}, \bibinfo{pages}{664} (\bibinfo{year}{2008}).

\bibitem[{\citenamefont{Koch et~al.}(2007)\citenamefont{Koch, Yu, Gambetta,
  Houck, Schuster, Majer, Blais, Devoret, Girvin, and Schoelkopf}}]{koch:2007a}
\bibinfo{author}{\bibfnamefont{J.}~\bibnamefont{Koch}},
  \bibinfo{author}{\bibfnamefont{T.~M.} \bibnamefont{Yu}},
  \bibinfo{author}{\bibfnamefont{J.}~\bibnamefont{Gambetta}},
  \bibinfo{author}{\bibfnamefont{A.~A.} \bibnamefont{Houck}},
  \bibinfo{author}{\bibfnamefont{D.~I.} \bibnamefont{Schuster}},
  \bibinfo{author}{\bibfnamefont{J.}~\bibnamefont{Majer}},
  \bibinfo{author}{\bibfnamefont{A.}~\bibnamefont{Blais}},
  \bibinfo{author}{\bibfnamefont{M.~H.} \bibnamefont{Devoret}},
  \bibinfo{author}{\bibfnamefont{S.~M.} \bibnamefont{Girvin}},
  \bibnamefont{and} \bibinfo{author}{\bibfnamefont{R.~J.}
  \bibnamefont{Schoelkopf}}, \bibinfo{journal}{Phys. Rev. A} \textbf{\bibinfo{volume}{76}},
  \bibinfo{eid}{042319} (\bibinfo{year}{2007}).

\bibitem[{\citenamefont{Schreier et~al.}(2008)\citenamefont{Schreier, Houck,
  Koch, Schuster, Johnson, Chow, Gambetta, Majer, Frunzio, Devoret
  et~al.}}]{schreier:2008a}
\bibinfo{author}{\bibfnamefont{J.~A.} \bibnamefont{Schreier}},
  \bibinfo{author}{\bibfnamefont{A.~A.} \bibnamefont{Houck}},
  \bibinfo{author}{\bibfnamefont{J.}~\bibnamefont{Koch}},
  \bibinfo{author}{\bibfnamefont{D.~I.} \bibnamefont{Schuster}},
  \bibinfo{author}{\bibfnamefont{B.~R.} \bibnamefont{Johnson}},
  \bibinfo{author}{\bibfnamefont{J.~M.} \bibnamefont{Chow}},
  \bibinfo{author}{\bibfnamefont{J.~M.} \bibnamefont{Gambetta}},
  \bibinfo{author}{\bibfnamefont{J.}~\bibnamefont{Majer}},
  \bibinfo{author}{\bibfnamefont{L.}~\bibnamefont{Frunzio}},
  \bibinfo{author}{\bibfnamefont{M.~H.} \bibnamefont{Devoret}},
  \bibnamefont{et~al.}, \bibinfo{journal}{Phys. Rev. B} \textbf{\bibinfo{volume}{77}}, \bibinfo{eid}{180502(R)} (\bibinfo{year}{2008}).

\bibitem[{\citenamefont{Yang and Han}(2005)}]{yang:2005a}
\bibinfo{author}{\bibfnamefont{C.-P.} \bibnamefont{Yang}} \bibnamefont{and}
  \bibinfo{author}{\bibfnamefont{S.}~\bibnamefont{Han}},
  \bibinfo{journal}{Phys. Rev. A}
  \textbf{\bibinfo{volume}{72}}, \bibinfo{eid}{032311} (\bibinfo{year}{2005}).

\bibitem[{\citenamefont{Lindstrom et~al.}(2007)\citenamefont{Lindstrom,
  Webster, Healey, Colclough, Muirhead, and Tzalenchuk}}]{lindstrom:2007a}
\bibinfo{author}{\bibfnamefont{T.}~\bibnamefont{Lindstrom}},
  \bibinfo{author}{\bibfnamefont{C.~H.} \bibnamefont{Webster}},
  \bibinfo{author}{\bibfnamefont{J.~E.} \bibnamefont{Healey}},
  \bibinfo{author}{\bibfnamefont{M.~S.} \bibnamefont{Colclough}},
  \bibinfo{author}{\bibfnamefont{C.~M.} \bibnamefont{Muirhead}},
  \bibnamefont{and} \bibinfo{author}{\bibfnamefont{A.~Y.}
  \bibnamefont{Tzalenchuk}}, \bibinfo{journal}{Supercon. Sci. Tech.} \textbf{\bibinfo{volume}{20}}, \bibinfo{pages}{814}
  (\bibinfo{year}{2007}).

\bibitem[{\citenamefont{A. A.~Abdumalikov Jr.
  et~al.}(2008)\citenamefont{A. A.~Abdumalikov Jr, Astafiev, Nakamura,
  Pashkin, and Tsai}}]{abdufarrukh-a.-abdumalikov:2008a}
\bibinfo{author}{\bibfnamefont{A. A.}~\bibnamefont{Abdumalikov, Jr.}},
  \bibinfo{author}{\bibfnamefont{O.}~\bibnamefont{Astafiev}},
  \bibinfo{author}{\bibfnamefont{Y.}~\bibnamefont{Nakamura}},
  \bibinfo{author}{\bibfnamefont{Y.~A.} \bibnamefont{Pashkin}},
  \bibnamefont{and} \bibinfo{author}{\bibfnamefont{J.~S.}~\bibnamefont{Tsai}},
  \bibinfo{journal}{Phys. Rev. B}
  \textbf{\bibinfo{volume}{78}}, \bibinfo{eid}{180502(R)} (\bibinfo{year}{2008}).

\bibitem{liu:2005a}
Y.-x. Liu, J.~Q. You, L.~F. Wei, C.~P. Sun, and F.~Nori, Phys. Rev. Lett.
  \textbf{95}, 087001 (2005).

\bibitem[{\citenamefont{Werner and Imamoglu}(1999)}]{werner:1999a}
\bibinfo{author}{\bibfnamefont{M.~J.} \bibnamefont{Werner}} \bibnamefont{and}
  \bibinfo{author}{\bibfnamefont{A.}~\bibnamefont{Imamoglu}}, \bibinfo{journal}{Phys. Rev. A}
  \textbf{\bibinfo{volume}{61}}, \bibinfo{pages}{011801(R)}
  (\bibinfo{year}{1999}).

\bibitem[{\citenamefont{Ledoux et~al.}(2005)\citenamefont{Ledoux, Daele, and
  Berghe}}]{ledoux:2005a}
\bibinfo{author}{\bibfnamefont{V.}~\bibnamefont{Ledoux}},
  \bibinfo{author}{\bibfnamefont{M.~V.} \bibnamefont{Daele}}, \bibnamefont{and}
  \bibinfo{author}{\bibfnamefont{G.~V.} \bibnamefont{Berghe}},
  \bibinfo{journal}{ACM Trans. Math. Softw.} \textbf{\bibinfo{volume}{31}},
  \bibinfo{pages}{532} (\bibinfo{year}{2005}).

\bibitem[{\citenamefont{Bourassa et~al.}(2009)\citenamefont{Bourassa, Gambetta,
 Brink, Schuster, Schoelkopf, Devoret and  Blais}}]{bourassa:2009a}
\bibinfo{author}{\bibfnamefont{J.}~\bibnamefont{Bourassa}},
  \bibinfo{author}{\bibfnamefont{J.~M.} \bibnamefont{Gambetta}},
  \bibinfo{author}{\bibfnamefont{M.}~\bibnamefont{Brink}},
  \bibinfo{author}{\bibfnamefont{D.~I}~\bibnamefont{Schuster}},
  \bibinfo{author}{\bibfnamefont{R.~J.}~\bibnamefont{Schoelkopf}},
  \bibinfo{author}{\bibfnamefont{M.~H}~\bibnamefont{Devoret}},
  \bibnamefont{and} \bibinfo{author}{\bibfnamefont{A.}~\bibnamefont{Blais}}
 \bibinfo{note}{(unpublished)}.

\bibitem{devoret:1997a}
M.~H. Devoret, \emph{Les Houches, Session LXIII, 1995}, edited by
  S.~Reynaud, E.~Giacobino and J.~Zinn-Justin (Elsevier Science, Amsterdam
  1997), p.~351.

\bibitem[{\citenamefont{Orlando et~al.}(1999)\citenamefont{Orlando, Mooij,
  Tian, van~der Wal, Levitov, Lloyd, and Mazo}}]{orlando:1999a}
\bibinfo{author}{\bibfnamefont{T.~P.} \bibnamefont{Orlando}},
  \bibinfo{author}{\bibfnamefont{J.~E.} \bibnamefont{Mooij}},
  \bibinfo{author}{\bibfnamefont{L.}~\bibnamefont{Tian}},
  \bibinfo{author}{\bibfnamefont{C.~H.} \bibnamefont{van~der Wal}},
  \bibinfo{author}{\bibfnamefont{L.~S.} \bibnamefont{Levitov}},
  \bibinfo{author}{\bibfnamefont{S.}~\bibnamefont{Lloyd}}, \bibnamefont{and}
  \bibinfo{author}{\bibfnamefont{J.~J.} \bibnamefont{Mazo}},
  \bibinfo{journal}{Phys. Rev. B} \textbf{\bibinfo{volume}{60}},
  \bibinfo{pages}{15398} (\bibinfo{year}{1999}).

\bibitem[{\citenamefont{Walls and Milburn}(1994)}]{walls:1994a}
\bibinfo{author}{\bibfnamefont{D.}~\bibnamefont{Walls}} \bibnamefont{and}
  \bibinfo{author}{\bibfnamefont{G.}~\bibnamefont{Milburn}},
  \emph{\bibinfo{title}{Quantum optics}} (\bibinfo{publisher}{Spinger-Verlag},
  \bibinfo{address}{Berlin}, \bibinfo{year}{1994}).

\bibitem[{\citenamefont{Blais et~al.}(2007)\citenamefont{Blais, Gambetta,
  Wallraff, Schuster, Girvin, Devoret, and Schoelkopf}}]{blais:2007a}
\bibinfo{author}{\bibfnamefont{A.}~\bibnamefont{Blais}},
  \bibinfo{author}{\bibfnamefont{J.}~\bibnamefont{Gambetta}},
  \bibinfo{author}{\bibfnamefont{A.}~\bibnamefont{Wallraff}},
  \bibinfo{author}{\bibfnamefont{D.~I.} \bibnamefont{Schuster}},
  \bibinfo{author}{\bibfnamefont{S.~M.} \bibnamefont{Girvin}},
  \bibinfo{author}{\bibfnamefont{M.~H.} \bibnamefont{Devoret}},
  \bibnamefont{and} \bibinfo{author}{\bibfnamefont{R.~J.}
  \bibnamefont{Schoelkopf}}, \bibinfo{journal}{Phys. Rev. A} \textbf{\bibinfo{volume}{75}},
  \bibinfo{eid}{032329} (\bibinfo{year}{2007}).
 
 \bibitem[{\citenamefont{Chiorescu et~al.}(2003)\citenamefont{Chiorescu,
  Nakamura, Harmans, and Mooij}}]{chiorescu:2003a}
\bibinfo{author}{\bibfnamefont{I.}~\bibnamefont{Chiorescu}},
  \bibinfo{author}{\bibfnamefont{Y.}~\bibnamefont{Nakamura}},
  \bibinfo{author}{\bibfnamefont{C.~J. P.~M.} \bibnamefont{Harmans}},
  \bibnamefont{and} \bibinfo{author}{\bibfnamefont{J.~E.} \bibnamefont{Mooij}},
  \bibinfo{journal}{Science} \textbf{\bibinfo{volume}{299}},
  \bibinfo{pages}{1869} (\bibinfo{year}{2003}).
  
 
 \bibitem{ciuti:2005a}
C.~Ciuti, G.~Bastard, and I.~Carusotto, Phys. Rev. B \textbf{72}, 115303
  (2005).

\bibitem{ciuti:2006a}
C.~Ciuti and I.~Carusotto, Phys. Rev. A \textbf{74}, 033811 (2006).
\bibitem[{\citenamefont{Knight et~al.}(1978)\citenamefont{Knight, Aharonov, and
  Hsieh}}]{knight:1978a}
\bibinfo{author}{\bibfnamefont{J.~M.} \bibnamefont{Knight}},
  \bibinfo{author}{\bibfnamefont{Y.}~\bibnamefont{Aharonov}}, \bibnamefont{and}
  \bibinfo{author}{\bibfnamefont{G.~T.~C.} \bibnamefont{Hsieh}},
  \bibinfo{journal}{Phys. Rev. A} \textbf{\bibinfo{volume}{17}},
  \bibinfo{pages}{1454} (\bibinfo{year}{1978}).

\bibitem[{\citenamefont{Larson}(2008)}]{larson:2008a}
\bibinfo{author}{\bibfnamefont{J.}~\bibnamefont{Larson}},
  \bibinfo{journal}{Phys. Rev. A}
  \textbf{\bibinfo{volume}{78}}, \bibinfo{eid}{033833} (\bibinfo{year}{2008}).
  
\bibitem{jahn:1937a}
H.~A. Jahn and E.~Teller, Proc. R. Soc. Lond. A \textbf{161}, 220 (1937).

\bibitem[{\citenamefont{Hines et~al.}(2004)\citenamefont{Hines, Dawson,
  McKenzie, and Milburn}}]{hines:2004a}
\bibinfo{author}{\bibfnamefont{A.~P.} \bibnamefont{Hines}},
  \bibinfo{author}{\bibfnamefont{C.~M.} \bibnamefont{Dawson}},
  \bibinfo{author}{\bibfnamefont{R.~H.} \bibnamefont{McKenzie}},
  \bibnamefont{and} \bibinfo{author}{\bibfnamefont{G.~J.}
  \bibnamefont{Milburn}}, \bibinfo{journal}{Phys. Rev. A}
  \textbf{\bibinfo{volume}{70}}, \bibinfo{pages}{022303}
  (\bibinfo{year}{2004}).

\bibitem{meaney:2009a}
C.~P. Meaney, T.~Duty, R.~H. McKenzie, and G.~J. Milburn, e-print arXiv:0903.2681.
\bibitem{gambetta:2006a}
J.~Gambetta, A.~Blais, D.~I. Schuster, A.~Wallraff, L.~Frunzio, J.~Majer, M.~H.
  Devoret, S.~M. Girvin and R.~J. Schoelkopf, Phys. Rev. A \textbf{74}, 042318 (2006).
  
\bibitem{schuster:2007a}
D.~I. Schuster, A.~A. Houck, J.~A. Schreier, A.~Wallraff, J.~M. Gambetta,
  A.~Blais, L.~Frunzio, J.~Majer, B.~Johnson, M.~H. Devoret, S.~M. Girvin, and
  R.~J. Schoelkopf, Nature (London) \textbf{445}, 515 (2007).
 
\bibitem[{\citenamefont{Simons}(2001)}]{simons:2001a}
\bibinfo{author}{\bibfnamefont{R.~N.} \bibnamefont{Simons}},
  \emph{\bibinfo{title}{Coplanar Waveguide Circuits, Components, and Systems}}
  (\bibinfo{publisher}{John Wiley \& Sons, Inc.}, \bibinfo{year}{2001}).

\bibitem{watanabe:1994a}
K.~Watanabe, K.~Yoshida, T.~Aoki and S.~Kohjiro, Jap. J. Appl.
  Phys. \textbf{33} 5708 (1994).

\bibitem[{\citenamefont{Burkard and Brito}(2005)}]{burkard:2005c}
\bibinfo{author}{\bibfnamefont{G.}~\bibnamefont{Burkard}} \bibnamefont{and}
  \bibinfo{author}{\bibfnamefont{F.}~\bibnamefont{Brito}},
  \bibinfo{journal}{Phys. Rev. B }
  \textbf{\bibinfo{volume}{72}}, \bibinfo{eid}{054528}
 (\bibinfo{year}{2005}).


\end{thebibliography}

\end{document}